\documentclass[useAMS,usenatbib]{mn2e}
\usepackage{graphicx}
\usepackage{amssymb}
\usepackage{amsmath}
\usepackage{bm}
\usepackage{epstopdf}

\newcommand{\lya}{Ly$\alpha$}
\newcommand{\lyb}{Ly$\beta$}
\newcommand{\nhi}{$N_{\rmn{HI}}$}
\newcommand{\ztrue}{z^{\star}_{N}}
\newcommand{\SumOne}{\mathcal{S}}
\newcommand{\SumTwo}{\mathcal{S}_{\Delta t}}

\newcommand{\sumiN}{\sum_{i=1}^{N}}

\title[\lya~absorbers \& the redshift drift]{\lya~absorbers in motion: consequences of gravitational lensing for the cosmological redshift drift experiment\thanks{Research undertaken as part of the Commonwealth Cosmology Initiative (CCI: www.thecci.org), an international collaboration supported by the Australian Research Council (ARC)} }

\author[M. Killedar \& G. F. Lewis]{
Madhura Killedar\thanks{E-mail: m.killedar@physics.usyd.edu.au (MK); \newline gfl@physics.usyd.edu.au (GFL)}
\& Geraint F. Lewis\footnotemark[2]\\\
Sydney Institute for Astronomy, School of Physics, A28, The University of Sydney, NSW, 2006, Australia
}

\begin{document}

\date{Accepted 2009 October 22. Received 2009 October 19; in original form 2009 September 11}

\pagerange{\pageref{firstpage}--\pageref{lastpage}} \pubyear{2009}

\maketitle

\label{firstpage}

\begin{abstract}
The evolution of the expansion rate of the Universe results in a drift in the redshift of distant sources over time. A measurement of this drift would provide us with a direct probe of expansion history. The Lyman $\alpha$ (\lya) forest has been recognized as the best candidate for this experiment, but the signal would be weak and it will take next generation large telescopes coupled with ultra-stable high resolution spectrographs to reach the cm~s$^{-1}$ resolution required. One source of noise that has not yet been assessed is the transverse motion of \lya~absorbers, which varies the gravitational potential in the line of sight and subsequently shifts the positions of background absorption lines. We examine the relationship between the pure cosmic signal and the observed redshift drift in the presence of moving \lya~clouds, particularly the collapsed structures associated with Lyman limit systems (LLSs) and damped Lyman $\alpha$ systems (DLAs). Surprisingly, the peculiar velocities and peculiar accelerations {\it both} enter the expression, although the acceleration term stands alone as an absolute error, whilst the velocity term appears as a fractional noise component. An estimate of the magnitude of the noise reassures us that the motion of the \lya~absorbers will not pose a threat to the detection of the signal.
\end{abstract}


\section{Introduction}
\label{intro}

The large-scale homogeneity and isotropy of the Universe is encompassed in Friedmann-Lema\^{i}tre-Robertson-Walker (FLRW) geometry (See \citet*{H06} for further details). One of the key aspects of this cosmology is the evolution of the scale factor, $R(t)$; modern cosmology aims to characterize this evolution, distinguish between competing cosmological models and predict the fate of the Universe. Observationally, this has been a rather difficult science, with so-called standard candles and standard rulers playing leading roles in indirect measurements. \citet{S62} recognized that a direct detection of the evolution of the expansion rate was, in theory, possible; just as the cosmological redshifts of extra-galactic objects in all directions are a consequence of isotropic expansion, a positive or negative drift in the redshift {\it of a single object} over time would indicate acceleration or deceleration with respect to the observer. The redshift drift is very small, but as we head towards an era of cm~s$^{-1}$ resolution spectroscopy, the prospect of detection becomes very real. In addition to identifying the appropriate instrumentation and spectroscopic techniques required, it is crucial to account for any systematic biases and consider if any source of noise may hinder the detection of the cosmic signal. Previous studies have covered some of this ground but the potential for the transverse motion of Lyman~$\alpha$ (\lya) clouds to hide this signal remains yet unexplored. In the present paper we aim to address this issue and estimate the magnitude of the bias or noise that is introduced.

In Section~\ref{background} we review the current cosmological paradigm along with relevant observational tests.
We re-derive the expression for the cosmic signal being sought in Section~\ref{cosmicsignal}.
In Section~\ref{movinglenseffect} we introduce the transverse moving lens effect.
We derive the expression for the observed redshift drift in the presence of \lya~clouds in Section~\ref{zdrift} and discuss the results in Section~\ref{conclusions}.

\section{Background}
\label{background}

The linear correlation between the recession velocities and luminosity distances of galaxies within a few Mpc was first discovered by \citet{H29}\footnote{There were large systematic errors in Hubble's distances and the scales probed were rather small; nevertheless, the result was fundamental to the acceptance of the expanding Universe hypothesis.}. This result, the first Hubble diagram, set the observational precedent for the notion of an expanding Universe, at a time when a static Universe was philosophically favoured.  More recently, two independent measurements of the luminosities of `standard candle' Type Ia supernovae (SNIa) by \citet{Ri98} and \citet{P99} have revealed that, assuming a metric theory of gravity, the Universe is currently undergoing a period of acceleration \citep{ST06}, and there is evidence for past deceleration beyond a transition redshift of $z \sim 0.5$ from Hubble Space Telescope (HST) observations of high redshift ($0.2 \la z \la 1.6$) SNIa \citep{Ri04}. 

Whilst deceleration is expected in a matter-only universe, recent acceleration requires a `dark energy' component. To uncover the underlying physical cause of the acceleration, it is necessary to map out the evolution of the scale factor; the functional form is characterised by the relative densities (in units of the critical density $\Omega_{0}$) of the various matter-energy components of the cosmological fluid $\Omega_{X,0}$, where $X$ may be one of $r$ (radiation), $M$ (matter), $DE$ (dark energy) or $k$ (curvature) and the subscript $0$ denotes the current values. The growing evidence points towards a flat universe that is currently dominated by a cosmological constant, $\Lambda$ (a `dark energy' component with an equation of state $w = -1$), a subdominant pressureless dark matter component and a small amount of baryonic matter. 

The expansion rate is normalised by its value today, i.e. the Hubble Constant: $H_{0} \equiv 100 h$ km~s$^{-1}$ Mpc$^{-1}$. The HST Key Project has combined several independent distance methods, establishing the most accurate measurement as $H_{0} = 72 \pm 8$~km~s$^{-1}$~Mpc$^{-1}$ \citep{F01}. This is in good agreement with the current best estimates for the values of the cosmological parameters:  $\Omega_{M,0}=0.27$, $\Omega_{\Lambda,0}=0.73$ \& $h=0.71$, based on observations of the Cosmic Microwave Background (CMB) radiation by the Wilkinson Microwave Anisotropy Probe 5-year data \citep[WMAP5;][]{K09}, SNIa \citep[][and references within]{K08}, and Baryonic Acoustic Oscillations \citep[BAOs;][]{P07} in the galaxy distribution. Each of the measurements individually provide constraints on various degenerate combinations of the parameters in question; the Hubble parameter constraints are model dependent.

Whilst precision cosmology converges upon a concordance in the parameters with complementary methods, the expansion remains an underlying assumption upon which observed phenomena, such as cosmological redshifts, are interpreted. Few direct tests are available to support this hypothesis. One observable consequence of expansion is time dilation by a factor of ($1+z$). \citet{W39} suggested the use of SNIa light curves as `cosmic clocks' and the results of high redshift SNIa observations have confirmed the expansion hypothesis and excluded other models \citep{Le96,G01,B08}. Another prediction is the decrease in the surface brightness (SB) of galaxies with a $(1+z)^{-4}$ redshift dependence, independent of all other cosmological parameters \citep{T30,T34,HT35}. The Tolman SB test has been performed in several studies, each confirming the expanding geometry. \citet*{PDdC96} and \citet{LS01} rule out the $(1+z)^{-1}$ dependence expected by the alternative non-expanding tired-light model at the $5\sigma$ and $10\sigma$ confidence level respectively. Furthermore, the temperature of the CMB is predicted to increase linearly with redshift: $T_{\rmn{CMB}}(z) = T_{0}(1+z)$; the temperature at the present epoch has been accurately determined to be $T_{0}=2.725\pm0.002 K$ \citep{M99}. Atomic fine structure transitions identified in quasar absorption spectra probe the CMB temperature at the absorption redshift \citep{BW68}. This method generally provides an upper limit to the temperature only, but the observations, which cover redshifts up to $z\la3$ are consistent with the standard cosmological model \citep*[e.g.][and references therein]{S94,S00,M02}. Alternatively, the CMB temperature can be estimated at the redshift of galaxy clusters by observing the thermal Sunyaev-Zel'dovich (SZ) effect \citep*{F78,Re80}; measurements using this method have been carried out for clusters at a range of redshifts $0.02<z<0.6$, again finding the redshift dependence to be consistent with an expansion hypothesis \citep{DP02,B02,L09}.

Redshift drift is a direct probe of the dynamics of expansion, as the Hubble parameter $H(z)$ is measured at particular redshifts. It allows a unique insight into the validity of our cosmological model, sometimes with complementary observations. For example, by characterising the redshift dependence of the drift, the experiment can constrain evolving dark energy models \citep*{C07} and map the equation of state \citep{L07}, although many-parameter dark energy models are harder to constrain \citep{BQ07}. If we live in a $\Lambda$CDM universe, the Chaplygin gas model \citep*{KMP01} and interacting dark energy model would be rejected at high confidence \citep{BQ07}, but the Lema\^{i}tre-Tolman-Bondi (LTB)  void models would be ruled out even earlier \citep{Q09}. Furthermore, when combined with luminosity distance data, the redshift drift experiment provides a test of the Copernican Principle \citep*{UCE08} and an independent measurement of spatial curvature \citep{NC99}.

Many of these tests rely on accurate measurements of the drift, and detection alone is only possible with next generation large telescopes. All systematic biases and sources of noise must, therefore, be identified and the threat they pose to the detection must be quantified. With this in mind, we examine one such threat: the changes to observed redshifts of objects caused by the transverse motion of \lya~clouds.

 
\section{The Cosmic Deceleration Signal}
\label{cosmicsignal}
 
The general form for the redshift drift dependence on redshift and the expansion rate was originally derived by \citet[][see eqn. 4A]{MV62}. We briefly re-derive it here.
 
 The Hubble parameter is defined in terms of the scale factor and expansion rate:
\[
	H(t) \equiv \frac{\dot{R}(t)}{R(t)}
\]
 The cosmological redshift of an emitter is related to the ratio of the scale factors at emitter and observer:
\[
	1+z = \frac{R(t_{0})}{R(t)}
\]
where $t_{0}$ is the time of observation and $t$ the time of emission.

As an aside, the cosmological fluid components appear in the redshift dependence of the Hubble parameter as derived from the cosmological field equations. The expression is given by:
\begin{equation*}
	H(z) =  H_{0}\sqrt{\Omega_{M,0}(1+z)^{3}+\Omega_{\Lambda,0}+\Omega_{k,0}(1+z)^{2}}
\end{equation*}
where we neglect radiation and assume dark energy to be in the form of a cosmological constant.

Consider now the redshift after some time interval, when the scale factors have evolved at both emitter and observer:
\begin{align}
	1+z + \Delta z 	&= \frac{R(t_{0}+\Delta t_{0})}{R(t+\Delta t)} \notag \\
				&\approx  \frac{R(t_{0}) + \dot{R}(t_{0})\Delta t_{0}}{R(t) + \dot{R}(t)\Delta t} \notag \\
				&= (1+z) \frac{1+H(t_{0})\Delta t_{0}}{1+H(t)\Delta t} \notag \\
				&= (1+z)(1+H_{0}\Delta t_{0} - H(z)\Delta t) \notag
\end{align}
The scale factors have been expanded to first order in $\Delta t$. The Hubble parameter at the time of observation is relabelled:  $H_{0} \equiv H(t_{0})$, whilst the Hubble parameter at emission is expressed in terms of the {\it redshift} of the emitter, rather than the {\it time} of emission i.e. $H(z)$ rather than $H(t)$. Subtracting to obtain the change in redshift:
\[
	\Delta z = (1+z)(H_{0}\Delta t_{0} - H(z)\Delta t).
\]
 If both emitter and observer are at rest in comoving coordinates, then time intervals in their respective frames are related by:
\[
 	\Delta t_{0} = \Delta t (1+z) 
\]
and thus we obtain the expression for the redshift drift:
 \begin{equation}
 	\Delta z = (1+z-E(z))H_{0}\Delta t_{0}
	\label{McVL}
 \end{equation}
where 
\[
	E(z) \equiv H(z)/H_{0}.
\]
The exact functional form is dependent on the equation of state and density of the cosmological fluid components, but an order of magnitude estimate:
\[
	\Delta z \approx 1 \times 10^{-10} h (1+z-E(z))\left( \frac{\Delta t_{0}}{yr} \right)
\]
using the density parameters determined from WMAP5+BAO+SNIa and evaluated, for example, at $z=3$, gives an estimated cosmic signal of $| \Delta z| \sim  4 \times 10^{-10}$, or $\Delta$v $\sim 3 \times 10^{-2}$ m~s$^{-1}$, per decade between epochs of observation.

\citet{S62} acknowledged that the required spectral resolution ($\sim$1 cm~s$^{-1}$ per decade cadence) was beyond the capabilities of the instrumentation of that generation. The redshift drift and its potential to constrain cosmological parameters has been studied within the contexts of various cosmological models \citep{ET75,Ru80,L81}, but without hope of detection.

\subsection{The New Challenge}
\label{newchallenge}
\citet{L98} identified the \lya~forest as the best candidate for a redshift drift experiment. The \lya~forest is a phenomenon observed in the spectra of quasars as a series of absorption lines blueward of the quasar \lya~emission line, each at a rest wavelength of $\lambda_{\rmn{Ly}\alpha} \equiv 1216 $\AA. The absorption lines are fingerprints of intervening neutral hydrogen clouds. Although absorbers may lie anywhere between the observer and the quasar, they are not detected or distinguishable below a certain redshift cutoff; the reasons are two-fold. Quasar spectra are usually observed by ground-based optical telescopes, but low-redshift absorption lines lie in the ultraviolet, so only absorbers at $z\ga1.7$ are detected\footnote{Although space-based UV spectrographs may detect the low-redshift \lya~forest, they would not have the stability required for such an experiment; the ultra-stable high resolution instrumentation is necessarily ground-based.}. Contamination by the \lyb~emission line ($\lambda_{\rmn{Ly}\beta} \equiv 1025 $\AA), and the associated \lyb~forest, causes confusion below: 
$1+z = (1+z_{\rmn{q}}) \lambda_{\rmn{Ly}\beta} / \lambda_{\rmn{Ly}\alpha}$. The redshift range thus probed with this experiment does not include the current acceleration phase. The \lya~forest supplies a large sample of spectral lines at high redshift, so while the linewidths are relatively large ($\sim20$ km~s$^{-1}$) and the cosmic signal weak, the sheer number density of lines per redshift bin should yield the necessary statistical accuracy. Loeb found that existing spectroscopic techniques, such as those employed in extra-solar planet searches, could produce a marginal detection.
 
This experiment has become one of the science drivers for next generation 30--60 m Extremely Large Telescopes (ELTs). The instrumental challenges for the COsmic Dynamics EXperiment (CODEX) spectrograph proposed for the European-ELT have been discussed by \citet{P05}, \citet{M07} and \citet{C07}; the latter finds CODEX capable of detecting the redshift drift in the \lya~forest over $\sim$10 yrs.  \citet[hereafter L08]{L08} conducted an extensive study into the feasibility of detection with a 42-m ELT, concluding that $\sim4000$ hrs of observing time over a 20 yr interval would reveal the redshift drift at a $\sim3\sigma$ significance.
 
But before the search for a weak signal such as this is undertaken, all possible systematic biases and sources of noise must be accounted for. Peculiar accelerations of the Lyman $\alpha$ clouds or their associated galaxies present perhaps the most obvious obstacle. The magnitude of the noise has been studied by \citet{P82}, \citet{L82}, \citet{L08} and \citet*{UBM08}; after some initial controversy, the accelerations were determined to have little impact on the signal.
 
The noise from galactic feedback and optical depth variations were evaluated by L08 and found to have a negligible effect. L08 also noted that the evolution of gravitational potential wells in the line of sight was a source of error not yet accounted for. We identify \lya~absorbers to be culprits; the positions of absorption lines may be shifted by foreground \lya~clouds, which are effectively weak lenses with non-zero transverse velocities.

  
\section{Transverse-Moving Lenses}
\label{movinglenseffect}
As light passes near a massive object it is gravitationally lensed, but if the lens gravitational potential field varies during the time of passage, the light experiences a net red- or blue-shift. This leads to the well known Rees-Sciama effect \citep{RS68} in the context of a collapsing galaxy cluster lens; a similar phenomenon occurs when a (non-evolving) lens moves across the line of sight.  \citet{BG83} were the first to describe how lenses with transverse relativistic peculiar velocities asymmetrically distort background emission and the general relativistic treatment was later developed by \citet{PB93}. 

This effect (hereafter the BG effect) has garnered attention largely in the context of galaxy clusters moving across the sky, inducing secondary anisotropies in the CMB brightness temperature \citep{BG83,B89,A98}\footnote{The induced CMB anisotropy derived by \citet{BG83} was incorrect by a factor of two. This was corrected by \citet{B89}.}. The influence of cluster lens transverse motion on weakly and strongly lensed background galaxies has also been investigated \citep{MB03}.
 
 The induced change in wavelength by the BG effect is proportional to the lens velocity and deflection angle. For a given lens speed, the magnitude of the change in wavelength is maximised if the lens is moving in the plane of the sky and towards (or away from) the image of the background source. For the purposes of this study, we need only consider this special case. We thus have:
 \begin{equation}
 	\Bigl\lvert\frac{\delta\lambda}{\lambda}\Bigr\rvert  =  \gamma \frac{\rmn{v}}{c} \hat{\alpha}
	\label{lenskick}
 \end{equation}
 where v is the lens speed, $\gamma$ is the Lorentz factor associated with the total lens velocity and $\hat{\alpha}$ is the deflection angle. This manifests as a change in the observed redshift:
\begin{align}
	\Bigl\lvert\Delta z\Bigr\rvert &= (1+z_{s})\Bigl\lvert\frac{\delta\lambda}{\lambda}\Bigr\rvert  \notag\\
			& = (1+z_{s}) \gamma \frac{\rmn{v}}{c} \hat{\alpha} \label{dzmovinglens}
\end{align}
where $z_{s}$ is the source (not lens) redshift. It is useful to consider the order of magnitude of the redshift:
\[
	\Bigl\lvert\Delta z\Bigr\rvert  \approx1.6\times10^{-8}(1+z_{s})\gamma\frac{\rmn{v}}{ ( 1000\mathrm{~km~s^{-1}} ) } \frac{\hat{\alpha}}{1"}
\]
A typical cluster lens ($\rmn{v} \sim1000$~km~s$^{-1}$, $\hat{\alpha}\sim30"$) can change the redshift of a source at $z\sim3$ by $\Delta z\sim2\times10^{-6}$ or $\Delta$v $\sim150$~m~s$^{-1}$, while a galaxy lens may produce an effect an order of magnitude smaller. 

In this paper, we interpret \lya~absorbers as a series of numerous moving lenses and consider the impact of all foreground absorbers on the observed wavelengths of absorption lines from more distant \lya~clouds. In the context of the redshift drift, however, we must determine the {\it differential} BG effect, i.e. how the shift in the observed wavelengths at the second epoch of observation compare to the shift at the first  epoch. Thus we will consider the transverse accelerations, $\rmn{a_{t}}$, as well.

  
\section{The effect of Lyman $\alpha$ Clouds on the observed redshift drift}
\label{zdrift}
 
We consider here the dynamic nature of \lya~clouds and how, as lenses, they can affect the redshift drift observed over some time interval. The expression we will derive below is parameterized by the cadence between observations, the number of \lya~absorbers, their peculiar velocities and accelerations as well as the deflection angle they induce.
 
\subsection{\lya~Clouds}
\label{lyaclouds}

\lya~clouds are typically distinguished by their neutral hydrogen column densities, \nhi, in units of cm$^{-2}$. Beyond log(\nhi) $\ga 17$, the optically thick nature of the absorbers results in a discontinuity at (blueward of) 912\AA~(rest-frame); these absorbers are classified as Lyman limit systems (LLS). Clouds with log(\nhi) $\ga 20$ exhibit damping wings about the absorption line profile leading to large \lya~equivalent widths; these are labelled damped Lyman $\alpha$ systems (DLAs). 
 
The exact nature of these absorbers is somewhat a mystery [see \citet{Ra98} for a review]. Low column density absorption occurs in filamentary or sheet-like structures of scale lengths of 0.1--1 Mpc; LLS absorbers are identified with the outer haloes of spiral and elliptical galaxies, whilst DLAs occur in spiral disks or the haloes of low-mass galaxies. The lensing effect of DLAs (and other types of systems such as those identified by metal lines) is significant enough to introduce a magnification bias, and push quasars over magnitude thresholds \citep{BL96,M05}. We must, therefore, consider whether the BG effect associated with these absorbers is significant.

The objects responsible for LLSs and DLAs are denser and therefore the location of higher peculiar accelerations, as recognized by \citet{L98}. L08 thus excluded them from their analysis, modelling only the low column density \lya~forest, and removing the portions of the spectra blocked out by DLAs from real spectra. {\it This does not mean that quasar spectra that contain LLSs and DLAs are excluded from the redshift drift experiment}. Indeed, each line of sight may include low-redshift LLSs and DLAs hidden outside the observed bandwidth; in fact, these play an important role in the present study.

\begin{figure*}
	\begin{center}
  		\includegraphics[width=150mm]{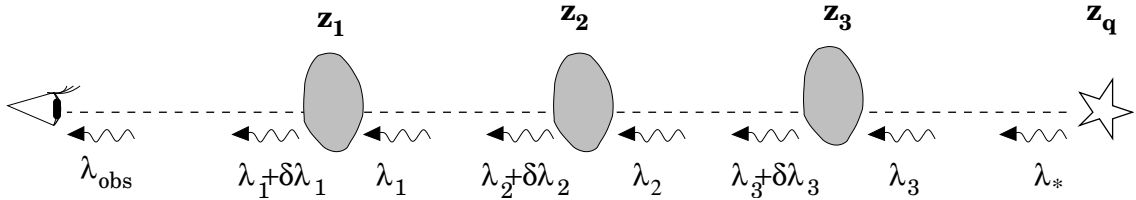}
	\end{center}
 	\vspace{3mm}
	\caption{Light from a distant quasar (right), emitted at rest wavelength $\lambda_{\ast}$ travels through many \lya~absorbers (three shown here), before reaching the observer (left) at the observed wavelength $\lambda_{obs}$. Each absorber, {\it i}, is also a lens with some non-zero transverse velocity (up or down the page) and thus each induces a change in the wavelength, $\delta\lambda_{i}$, in the frame of the observer. A photon is cosmologically redshifted as it travels between two adjacent clouds (see Eqn~\ref{betweenclouds}).}
 	\vspace{5mm}
	\label{figClouds}
\end{figure*}

\subsection{Velocities of Lenses}
\label{velocities}
Previous studies have provided insight into the peculiar motions of \lya~clouds. \citet{Ra05} measured the velocity shear between pairs of absorbers common to multiple images of gravitationally lensed quasars; their lines of sight are close together. Their results indicate that while the large-scale motions correspond to the Hubble flow, there is tentative evidence for the clouds undergoing gravitational collapse. L08 used hydrodynamic simulations of the intergalactic medium (IGM) to determine the peculiar motions of \lya~absorbers. They find that the distribution of peculiar velocities peak at v$_{\rmn{pec}} \sim $190--200 km~s$^{-1}$ for gas at $2\la z\la 4$ with an approximate dispersion of $\sigma_{\rmn{v_{pec}}} \sim$ 80 km~s$^{-1}$.

A single high-velocity cloud that is present in the line of sight, with absorption taking place at an impact parameter of 1~kpc from a $10^{11}M_{\odot}$ galactic halo (v $\sim 300$ km~s$^{-1}$, $\hat{\alpha}\sim4"$),  could alter positions of background absorption lines at redshift $z$ by as much as $\Delta z \sim 2 \times 10^{-8} (1+z)$. This is indicative of the BG effect as can be expected from the high column density absorbers, such as LLSs and some DLAs. 

The lower column density absorbers are associated with filamentary structures, although overdense knots will move along these, as seen in cosmological N-body simulations. In this context, equation~\ref{lenskick}, which strictly assumes a compact lens, would require an additional term to compensate for the extended shapes of the overdensities, rendering the BG effect less pronounced. The BG effect here may also  arise from galaxy clusters that lie at the ends of the associated filaments. A moderate sized cluster ($10^{13}M_{\odot}$) at a distance of 1~Mpc and with reasonable velocity (v $\sim 300$ km~s$^{-1}$) could potentially alter the observed redshifts by $\Delta z \sim 2 \times 10^{-9} (1+z)$. To ascertain whether this poses a threat to the detection of the cosmic signal, we must include the effect in the derivation of the observed redshift drift.

\subsection{Time Interval between Observation Epochs}
\label{timeinterval}
The duration of the interval between observations of quasar spectra has by far the largest impact on the chance to detect the cosmic signal. The chosen cadence must clearly be large enough to allow for some appreciable evolution in the scale factor (over all redshifts $2 \la z \la 5$). We will, however, still have $H_{0}\Delta t_{0} \ll 1$ and $\Delta z \ll z$. By combining the signal from $\sim100$ uncorrelated quasars, the signal may be detected over only $\Delta t_{0} \sim$ 10--20 yrs \citep{L98,C07,L08}. These intervals are proposed in the context of spectrographs mounted on 30--60 m ELTs proposed for the near future.

\subsection{The Observed Redshift Drift}
\label{observedzdrift}
Here we derive the observed redshift drift over some time interval and compare this to the drift resulting purely due to cosmic deceleration. Figure~\ref{figClouds} shows how the wavelength of a photon varies as it travels from a distant source to an observer being lensed by moving \lya~clouds along the way. We have $0<z_{i}<z_{j}<z_{q}$ for $i<j$, although we are not particularly concerned with the redshift of the source quasar nor the wavelength of the photon upon emission, $\lambda_{\ast}$. We are, instead, interested in the redshift of an absorption line with rest wavelength $\lambda_{i}$, observed at $\lambda_{obs}$.
 
Each photon (or absorption line) is red- or blue-shifted in the frame of a stationary observer upon passing a \lya~cloud with some non-zero transverse velocity; the photon enters the cloud at $\lambda$ and exits at $\lambda+\delta\lambda$, where the small shift is given by Eqn.~\ref{lenskick}. Absorption occurs in the rest frame of the cloud, therefore, in the stationary frame (though at the same redshift), the absorption line is shifted in the same manner. 
For brevity, we denote:
\[
	\lambda' \equiv \lambda+\delta\lambda.
\]
When faced with the situation where the observed wavelength of any emission can be altered by some means other than expansion, we must distinguish between the {\it true} redshift, $z^{\star}$, and that which is observed, $z^{obs}$. The true redshift is given by:
\[
	1+z^{\star} = \frac{R(t_{0})}{R(t)}
\]
while the observed redshift is defined as:
\[
	1+z^{obs}= \frac{\lambda_{obs}}{\lambda}.
\]
As the photon travels from cloud $j$ (or the source quasar) to the next cloud $j-1$ (or observer) it is redshifted as a result of the difference in scale factors at the two clouds.
Thus the (true) redshifts of the two clouds are related by:
 \begin{equation}
 	\frac{1+z^{\star}_{j}}{1+z^{\star}_{j-1}} = \frac{\lambda_{j-1}}{\lambda_{j}'}.
	\label{betweenclouds}
 \end{equation}
We recognize that the true redshift of the $N^{\rmn{th}}$ cloud is the product of the ratios of scale factors at consecutive pairs of foreground clouds, and so:
 \begin{align}
 	1+\ztrue	&= \prod_{i=1}^{N}\frac{\lambda_{i-1}}{\lambda_{i}'} \notag\\
			&= \frac{\lambda_{obs}}{\lambda_{N}}\prod_{i=1}^{N}\frac{\lambda_{i}}{\lambda_{i}'} \notag\\
			&\approx (1+z^{obs}_{N})\left[ 1 - \sum_{i=1}^{N}\frac{\delta\lambda_{i}}{\lambda_{i}}     \right] 
\label{withcloudsbefore}
 \end{align}
since $\delta\lambda\ll\lambda$. We have defined $\lambda_{0}\equiv\lambda_{\rmn{obs}}$. After some time interval $\Delta t_{0}$ (in the observer rest frame), the true redshift has evolved to be:
 \begin{align}
 	 1+\ztrue+\Delta \ztrue
	 	&\approx \frac{R_{0}+\dot{R}_{0}\Delta t_{0}}{R_{N}+\dot{R}_{N}\Delta t_{N}}  \notag\\
	 	&= (1+\ztrue)\frac{1+H_{0}\Delta t_{0}}{1+H(\ztrue)\Delta t_{N}}\notag \\
	 	&\approx 1+\ztrue + H_{0}\Delta t_{0} (1 +  \ztrue - E(\ztrue))
\label{withcloudsafter}
 \end{align}
 where $\Delta t_{N}$ denotes the time passed in the rest frame of the $N^{\rmn{th}}$ cloud:
\[
 	\Delta t_{0} = \Delta t_{N} (1+\ztrue).
\]
We observe a drifted redshift, analogous to equation~\ref{withcloudsbefore}:
\begin{equation}
 	 1+\ztrue+\Delta \ztrue = (1+z^{obs}_{N}+\Delta z^{obs}_{N})(1-\mathcal{S}) - (1+z^{obs}_{N})\mathcal{S}_{\Delta t}
	 \label{withcloudsafter2}
\end{equation}
where the summations are denoted thus:
\[
	\SumOne \equiv \sumiN\frac{\delta\lambda_{i}}{\lambda_{i}}  
\]
\[
	\SumTwo \equiv \Delta t_{0}\sumiN\frac{\rmn{a}_{\rmn{t},i}}{c} \hat{\alpha}_{i}
\]
with $\rmn{a}_{\rmn{t},i}$ the tangential peculiar {\it acceleration} of the $i{\rmn{th}}$ \lya~cloud. Any change in the deflection angle is dependent on the mass profile of the cloud and vanishes in the case of a singular isothermal sphere (SIS). In any case, the impact parameter changes by only a fraction of a parsec so we neglect this term.

We subtract the observations at the two epochs to find the observed drift rate i.e. Equate equations  \ref{withcloudsafter} \& \ref{withcloudsafter2} and subtract equation \ref{withcloudsbefore}:
 \begin{equation}
 	  H_{0}\Delta t (1 +  \ztrue - E(\ztrue)) = \Delta z^{obs}_{N}(1-\SumOne) - (1+z^{obs}_{N})\SumTwo.
	 \label{zdriftwithaccel}
 \end{equation}
As a sanity check, if we neglect the effect of all clouds, we recover the McVittie expression (Eqn.~\ref{McVL}).

\subsection{The magnitude of the noise}
\label{magnitudeofnoise}

Equation~\ref{zdriftwithaccel} describes how the observed redshift drift after a time interval $\Delta t_{0}$ differs from the drift expected purely due to the evolution of the expansion rate. The peculiar velocities of each absorber lens are contained within the summation $\SumOne$; even in the absence of peculiar acceleration, the velocities will affect the observed magnitude of the redshift drift by entering the expression as a fractional offset. The accelerations are contained within the summation $\SumTwo$, which stands alone as an absolute error. Below, we estimate the order of magnitude of these effects.

Each term in $\SumOne$~and $\SumTwo$~may be positive or negative depending on the direction of motion and acceleration of the \lya~cloud. Thus the uncertainties are in the form of noise, rather than a systematic bias; they are the result of a random walk. We would expect the noise to increase with absorption line redshift as there would be more lenses in the foreground.

The denser absorbers, such as LLSs, may produce deflection angles of at most a few arcseconds, and at speeds of $v\sim300$~km~s$^{-1}$ would dominate $\SumOne$ with $\Delta\lambda/\lambda\sim10^{-8}$, although we are unlikely to find more than one or two in a single quasar spectrum.  Even if we coupled the high column density absorbers with the expected large number of low column density absorbers ($ \sim1000$~per line of sight), and coincidentally aligned the directions of motions, we would still find $\bigl\lvert\SumOne \bigr\rvert \ll 1$. If $\SumOne$ was non-negligible, the net result would depend on how the lines are binned for the experiment. If the binned absorption lines come from one spectrum only, then $\SumOne$ for each line will be highly correlated, and a strong bias will be present. However, if multiple spectra are observed for the same experiment, which is likely to be the case, the value of $\SumOne$ for each spectrum would be independent, even for a particular redshift bin, and thus the effect would still average out.

L08 also plotted the narrow distribution of peculiar acceleration of the IGM within hydrodynamic simulations, establishing a tight peak at $a_{\rmn{pec}}\sim10^{-11}$~cm~s$^{-2}$. Again, coupling this acceleration with a large deflection angle of a few arcseconds, a large number density ($\sim1000$) and a time interval of 20 yrs between epochs of observation, we find $\SumTwo \sim 10^{-15}$. Though this appears as an absolute error in the measurement, the redshift drift expected is $\Delta z \sim10^{-10}$, many orders of magnitude larger. Again, binning within a single spectrum can amplify this offset, but the use of multiple spectra renders the offset small again, leaving the signal unaffected.

\section{Concluding remarks}
\label{conclusions}
 
 The redshift drift experiment has been recognized as a direct test of the cosmic
expansion, and carries with it the potential to independently constrain cosmological
parameters and distinguish between evolving dark energy models. The proposed next
generation of ELTs and ultra-stable high resolution spectrographs make the detection of
the cosmic signal possible in the near future. However, as the signal is weak, it is
necessary to account for all possible systematic biases and sources of noise. One
source of noise that has, until now, remained unexplored is the variation in the gravitational field due to the transverse motion of \lya~absorbers. In this paper, we have examined
this effect, and characterised and quantified the noise introduced to the signal. To
summarise, we find that the peculiar velocities and accelerations of the \lya~clouds both enter the expression relating the observed redshift to the cosmic signal as
noise terms; the former introduces a fractional offset, the latter, absolute. The
evolution of the gravitational potential wells in the line of sight due to transverse
motion of the \lya~absorbers will not, in fact, introduce significant noise
or bias into the detection of the cosmic signal.

   We consider now whether this is the final word on evolving gravitational fields. \lya~absorbers were the most obvious source of foreground interference; they, by definition, intersect the line of sight. Galactic haloes at impact parameters of 1--20~kpc are thought to produce the rarer absorption effects, such as LLSs and DLAs; these have been the extreme cases studied in this work. Estimates of the frequency and redshift distribution of these objects are hindered by reddening, which can dim quasars and exclude them from flux-limited samples, although the magnification bias can counteract this effect. Both effects can be expected, particularly if the absorbers are massive galaxy haloes at low impact parameters. The resulting evolution of the gravitational field within the filaments and sheets associated with the low-density \lya~forest is difficult to study analytically. The filaments feed into more massive structures, such as galaxy clusters, with absorption taking place at impact parameters of $\sim$1~Mpc. The transverse peculiar motion of such objects will have a similar `moving-lens' effect, the rarity of which has not been assessed. Their presence may not be noticed in the quasar spectra, though perhaps in direct imaging (at least at low redshift). Further studies using dark matter simulations will probe the expected motions of all mass scales in the vicinity of the line of sight to distant quasars, allowing the determination of the full influence of the BG effect in future redshift drift experiments; this will form the basis of future contributions.


\section*{Acknowledgments}
The authors would like to thank the anonymous referee for the helpful comments and suggestions. They acknowledge support from ARC Discovery Project DP0665574. M.K. is also supported by the University of Sydney Faculty of Science Postgraduate Award and would like to thank Richard Lane for his feedback.



\label{lastpage}
\end{document}